\documentclass{chjaa}                  

\newcommand{\ce}{\ifmmode {\cal E} \else ${\cal E}$\ \fi}
\newcommand{\kms}{\ifmmode {\rm km\ s}^{-1} \else km s$^{-1}$\ \fi}
\newcommand{\ergs}{\ifmmode {\rm erg\ s}^{-1} \else erg s$^{-1}$\ \fi}
\newcommand{\tes}{\ifmmode \tau_{\rm es} \else $\tau_{\rm es}$\ \fi}
\newcommand{\tk}{\ifmmode \tau_{\rm K} \else $\tau_{\rm K}$\ \fi}
\newcommand{\vfwhm}{\ifmmode V_{\mbox{\tiny FWHM}} \else
            $V_{\mbox{\tiny FWHM}}$\fi}
\newcommand{\msun}{\ifmmode M_{\odot} \else $M_{\odot}$\ \fi}
\newcommand{\afe}{\ifmmode {\mathcal A_{\rm Fe}} \else${\mathcal A_{\rm Fe}}$\ \fi}

\newcommand{\feii}{Fe {\sc ii}\ }
\newcommand{\mgii}{Mg {\sc ii}\ }
\newcommand{\civ}{C {\sc iv}\ }
\newcommand{\lb}{\ifmmode L_{\rm Bol} \else $L_{\rm Bol}$\ \fi}
\newcommand{\ledd}{\ifmmode L_{\rm Edd} \else $L_{\rm Edd}$\ \fi}
\newcommand{\lx}{\ifmmode L_{\rm 2-10keV} \else  $L_{\rm 2-10keV}$\ \fi}
\newcommand{\hb}{\ifmmode H\beta \else H$\beta$\ \fi}
\newcommand{\ha}{\ifmmode H\alpha \else H$\alpha$\ \fi}
\newcommand{\oiii}{[O {\sc iii}]\ }
\newcommand{\oii}{[O {\sc ii}]\ }
\newcommand{\sii}{[S {\sc ii}]\ }
\newcommand{\mbh}{\ifmmode M_{\rm BH}  \else $M_{\rm BH}$\ \fi}
\newcommand{\lv}{\ifmmode \lambda L_{\lambda}(5100\AA) \else $\lambda L_{\lambda}(5100\AA)$\ \fi}
\newcommand{\mdot}{\ifmmode \dot{m} \else \dot{m} \fi }
\newcommand{\llog}{\ifmmode {\rm log} \else {\rm log} \fi }

\usepackage{graphicx,times}             

\begin{document}

   \title{The radio luminosity, black hole mass and Eddington ratio for quasars from the Sloan Digital Sky Survey \footnotetext{$*$ Supported by the National Natural Science
Foundation of China.} }
   \setcounter{page}{1}           

   \author{Wei-Hao. Bian
      \inst{1,2}
   \and Yan-Mei. Chen
      \inst{1}
   \and Chen. Hu
      \inst{3}
         \and kai. Huang
      \inst{2}
   \and Yan. Xu
      \inst{2}
      }

   \institute{Key Laboratory for Particle Astrophysics, Institute of High
Energy Physics, Chinese Academy of Sciences, Beijing 100039, China
             \email{bianwh@ihep.ac.cn}
        \and
             Department of Physics and Institute of Theoretical
Physics, Nanjing Normal University, Nanjing 210097, China\\
        \and
             National Astronomical Observatories, Chinese Academy of Sciences,
             Beijing 100012, China\\
          }

   \date{}

   \abstract{
We investigate the $\mbh- \sigma_*$ relation for radio-loud
quasars with redshift $z<0.83$ in Data Release 3 of the Sloan
Digital Sky Survey (SDSS). The sample consists of 3772 quasars
with better model of H$\beta$ and \oiii lines and available radio
luminosity, including 306 radio-loud quasars, 3466 radio-quiet
quasars with measured radio luminosity or upper-limit of radio
luminosity (181 radio-quiet quasars with measured radio
luminosity). The virial supermassive black hole mass (\mbh) is
calculated from the broad \hb line, the host stellar velocity
dispersion ($\sigma_*$) is traced by the core \oiii gaseous
velocity dispersion, and the radio luminosity and the radio
loudness are derived from the FIRST catalog. Our results are
follows: (1) For radio-quiet quasars, we confirm that there is no
obvious deviation from the $\mbh- \sigma_*$ relation defined in
inactive galaxies when \mbh uncertainties and luminosity bias are
concerned. (2) We find that radio-loud quasars deviate much from
the $\mbh- \sigma_*$ relation respect to that for radio-quiet
quasars. This deviation is only partly due to the possible
cosmology evolution of the $\mbh- \sigma_*$ relation and the
luminosity bias. (3) The radio luminosity is proportional to
$\mbh^{1.28^{+0.23}_{-0.16}}(\lb/\ledd)^{1.29^{+0.31}_{-0.24}}$
for radio-quiet quasars and
$\mbh^{3.10^{+0.60}_{-0.70}}(\lb/\ledd)^{4.18^{+1.40}_{-1.10}}$
for radio-loud quasars. The weaker correlation of the radio
luminosity dependence upon the mass and the Eddington ratio for
radio-loud quasars shows that other physical effects would account
for their radio luminosities, such as the black hole spin.
   }

   \authorrunning{W. Bian, et al. }            
   \titlerunning{Radio-loud and radio-quiet quasars from SDSS}  

   \maketitle

\section{INTRODUCTION}
The relation between the supermassive black hole (SMBH) mass and
the host stellar velocity dispersion (hereafter $\mbh- \sigma_*$
relation) is one of the most important results in the study of
supermassive black holes (SMBHs) in these decades, implying the
intimate correlation between the SMBHs and their host galaxies
(e.g. Gebhardt et al. 2000; Ferrarese \& Merrit 2000; Tremaine et
al. 2002; Lauer et al. 2007). This correlation would provide
strong constraints for the evolution of active galactic nuclei
(AGNs) if we know AGNs follow this relation or not. However it is
still under debate for different kind of AGNs, such as radio-loud
AGNs, narrow-line Seyfert 1 galaxies, intermediate supermassive
black hole, et al. (e.g., Nelson 2001; Boroson 2003; Shield et al.
2003; Bian \& Zhao 2004; Grupe \& Mathur 2004; Bonning et al.
2005; Greene \& Ho 2006; Woo et al. 2006; Zhou et al. 2006;
Salviander et al. 2007; Komossa \& Xu 2007; Shen et al. 2008). In
order to study this relation for AGNs, we should calculate \mbh
and $\sigma_*$ as accurately as possible.

The width of the broad emission line (e.g., \hb, \ha, \mgii, \civ)
can be used to trace virial velocity of the clouds in broad line
regions (BLRs) when the line contribution from narrow-line regions
(NLRs) is reasonably removed, and the reverberation mapping method
or the empirical luminosity-size relation can be used to calculate
the BLRs size (e.g., Kaspi et al. 2000; McLure \& Dunlop 2004;
Bian \& Zhao 2004; Peterson et al., 2004; Greene \& Ho 2005b). The
gas velocity dispersion of the narrow lines (e.g., \oiii, \oii,
\sii) from NLRs are usually used to trace the host stellar
velocity dispersion (e.g., Nelson \& Whittle 1996; Greene \& Ho
2005a). We also can directly measure the host velocity dispersion
from AGNs host spectra (e.g., Kauffmann et al. 2003; Heckman et
al. 2004; Greene \& Ho 2005a; Bian et al. 2006). The larger number
of quasars found in the Sloan Digital Sky Survey (SDSS) provides
the possibility to tackle the $\mbh- \sigma_*$ relation in
radio-loud quasars. (e.g. Bian \& Zhao 2004; Salviander et al.
2007).

The dichotomy of the radio loudness in quasars is a long-time
question since the discovery of quasars (Sandage 1965;
Strittmatter et al. 1980; Kellermann et al. 1989). The radio
luminosity is assumed coming from the relativistic electrons
powered by a jet, which is intimately connected with the SMBH
(e.g., Begelman et al. 1984; Blundell \& Beasley 1998). For
scale-free jet physics and accretion theories, the radio
luminosity is related to the central engines, such the SMBH mass,
the SMBH spin, the Eddington ratio, et al. (Heinz \& Sunyaev
2003). For radio-loud or radio-quiet quasars, the dependence of
the radio loudness/luminosity upon the SMBH mass/Eddington ratio
is discussed by many peoples, some support it and some against it.
(e.g., Franceschini et al. 1998; Laor 2000; Lacy et al. 2001; Ho
2002; Woo \& Urry 2002; McLure \& Jarvis 2004; Wang et al. 2004;
Greene et al. 2006; Liu et al. 2006; Sikora 2007; Panessa et al.
2007). Laor (2003) gave some comments on the origin of AGNs radio
loudness and discussed the some error SMBH mass estimation for
radio-loud AGNs in the literature, which is mainly due to optical
spectra with low signal-to-noise ratios, no correction of H$\beta$
contribution from narrow line regions (NLRs).

In this paper, we use larger number of quasars with redshifts
$z<0.83$ in SDSS Data Release 3 (DR3; see Abazajian et al. 2005)
to investigate the $\mbh- \sigma_*$ relation and the radio
luminosity dependence on the SMBH mass and the Eddington ratio for
radio-loud and radio-quiet quasars. In \S 2, we briefly introduce
the SDSS quasars Data Release 3 catalog of Schneider et al.
(2005). \S 3 is the data analysis. \S 4 introduces the methods to
calculate the SMBH masses and the Eddington ratios. Our results
and discussions of $\mbh-\sigma_*$ relation and the origin of
radio luminosity are given in \S 5 and \S 6, respectively. The
last section is our conclusions. All of the cosmological
calculations in this paper assume $\rm H_{\rm 0}=70 \kms \rm
~Mpc^{-1}$, $\Omega_{\rm M}=0.3$, $\Omega_{\Lambda} = 0.7$.

\section{Sample and data analysis}
The sample used in this paper is selected from the SDSS quasars
Catalog III, which covers a spectroscopic area of 1360 sq. deg.,
about $40\%$ of the proposed SDSS survey area (Schneider et al.
2003). This catalog consists of 46,420 quasars in SDSS DR3 with
$M_{i}<-22$. The catalog also contains radio emission properties
from Faint Images of the Radio Sky at Twenty-cm (FIRST) survey
within 2.0" of the quasars position (see Col. 17 in their Table
1).

SDSS optical spectra cover the wavelength range 3800-9200 \AA\
with a resolution of $1800 < R < 2100$. In order to calculate SMBH
mass from the broad \hb line and the host stellar velocity
dispersion from the narrow \oiii line, we just consider the
quasars with redshifts less than 0.83, which consists of 9753
quasars. Because whether the SMBH mass from \mgii linewidth is
consistent with that from \hb line width is still a complex
question (e.g., Salviander et al. 2007), here we don't consider
using \mgii linewidth to calculate the SMBH mass.

The radio luminosity at 5GHz is calculated from the peak flux
density listed in Col. 17 in Table 2 (Schneider et al. 2003),
considering the spectral index of $\alpha=0.5$, where $f_{\nu}
\propto \nu ^{-\alpha}$. The radio loudness $R$ is calculated
from: $R=f_{\rm 5GHz}/f_{\rm B}$, where $f_{\rm 5GHz}$ and $f_{\rm
B}$ are the rest-frame flux density at 5 GHz and 4400\AA,
considering $k$ correction. $R=10$ is commonly used to define
radio-loud quasars and radio-quiet quasars (e.g., McLure \& Jarvis
2004), as well as the radio luminosity at 5 GHz (e.g., Lacy 2001).

For 9573 quasars with $z<0.83$ from SDSS DR3, 914 quasars are
detected by FIRST, 7846 quasars are under the FIRST flux limit,
and 993 quasars are not in the region covered by FIRST. For these
objects with non-detection in FIRST, we only have the upper-limits
of the radio luminosity and the radio loudness. 598 quasars with
detection in FIRST and $R\geq10$ are classified as radio-loud
quasars. 316 quasars with detection in FIRST and  $R<10$ are
classified as radio-quiet quasars. 5712 quasars with non-detection
in FIRST and $R<10$ are classified as radio-quiet quasars but with
upper-limits of $R$ and the radio luminosity.

\begin{figure}
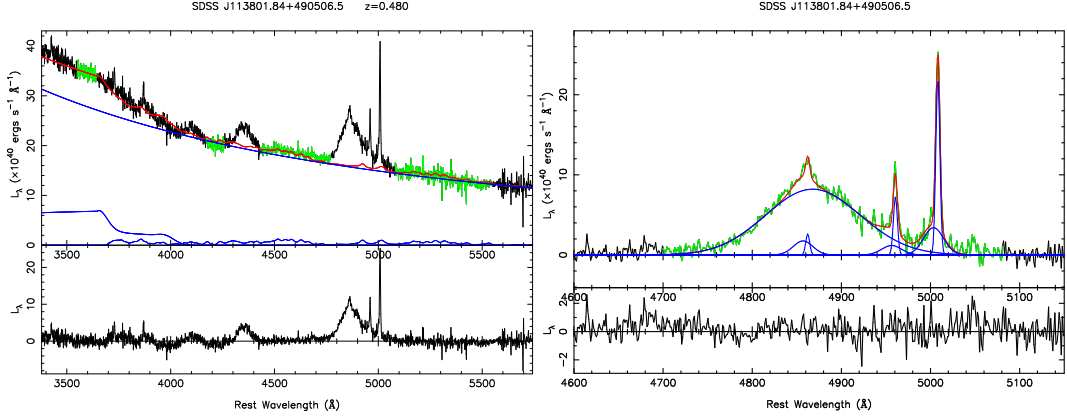

\begin{center}
\includegraphics[height=7cm,angle=-90]{f1a.eps}
\includegraphics[height=7cm,angle=-90]{f1b.eps}
\caption{Sample of SDSS spectrum measurement for SDSS
J113801.84+490506.5. In the top panel, the black curve is the
observed spectrum, the red line is the sum of the power-law
continuum, the Balmer continuum and \feii multiples (blue curves).
The green ranges are our fitting windows. The bottom panel is the
multi-Gaussian fit for \hb and \oiii lines. The red line is the
sum of all multi-Gaussian (blue curves). The green curve is our
fitting range of the pure \hb and \oiii emissions after the
subtraction of the power-law continuum, the Balmer continuum and
Fe multiples.}
\end{center}
\end{figure}

As we know, NLRs can contribute \hb emission in the total \hb
profile; \oiii usually shows non-symmetric profile and its
narrow/core component can trace the stellar velocity better (e.g.,
Greene \& Ho 2005a); optical and ultraviolet \feii multiples are
often presented in quasars spectra; Balmer continuum is required
because of the existence of strong Balmer emission lines,
therefore, we use following steps to do the SDSS spectral
measurements.

(1) First, we do the Galactic extinction in the observed spectra
by using the extinction law of Cardelli, Clayton \& Mathis (1989)
(IR band) and O'Donnell (1994) (optical band), then the spectra
are transformed into the rest frame defined by the redshifts given
in their FITS headers.

(2) The optical and ultraviolet \feii template from the prototype
NLS1 I ZW 1 is used to subtract the \feii emission from the
spectra (Boroson \& Green 1992; Vestergaard \& Wilkes 2001). The I
ZW 1 template is broadened by convolving with a Gaussian of
various linewidths and scaled by multiplying a factor. A power-law
continuum and the Balmer continuum are added in the fitting. We
calculate the Balmer continuum following Grandi (1982) and also
add the high order Balmer lines at the red side of the Balmer edge
using the result in Storey \& Hummer(1995). The best subtraction
of the \feii, power-law and Balmer continuum is found when
$\chi^2$ minimized in the fitting windows: 3550-3645, 4170-4260,
4430-4770, 5080-5550, 6050-6200, 6890-7010\AA\ (see a sample fit
in the top panel of Figure 1). The monochromatic luminosity at
5100\AA (\lv) is calculate from the power-law continuum.

(3) Two sets of two-Gaussian are used to model \oiii$\lambda
\lambda 4959, 5007$ lines. Three-Gaussian is used to model \hb
line. For the doublet \oiii $\lambda\lambda$4959,5007, we take the
same linewidth for each component, and fix the flux ratio of
\oiii$\lambda$4959 to \oiii $\lambda$5007 to be 1:3. Two
components of \hb (supposed from NLRs) are set to have the same
linewidth of each component of \oiii $\lambda$5007 and their flux
are constrained to be less than 1/2 of each component of \oiii
$\lambda 5007$. The linewidth of the broad component of \hb is
used to trace the virial velocity around central SMBH (see a
sample fit in the bottom panel of Figure 1).

From above spectral measurement, we obtain the full width at half
maximum (FWHM) of the broad \hb line and the narrow/core \oiii
line ($\rm FWHM_{\rm H\beta}$, $\rm FWHM^{n}_{\rm [O III]}$), the
monochromatic luminosity at 5100\AA (\lv), the total \hb
luminosity ($L_{\rm H\beta}$), as well as the radio luminosity and
the radio loudness for SDSS DR3 quasars with $z<0.83$.

Objects without the \hb or \oiii lines are eliminated. In order to
obtain the reliable spectra fit, we carefully select objects for
analysis. The line equivalent width (EW) can show line
signal-to-noise ratios. The error of EW can be regard as a tracer
to show the fitting goodness. Because the \hb is usually strong,
we don't constrain EW of \hb line, only constrain the error of EW
for \hb line. We select objects by the criterions of EW of \oiii
larger than 1.5, the errors of EWs of \hb and \oiii$\lambda
\lambda 4959, 5007$ less than $100\%$. It leads to 367 radio-loud
quasars, 3677 radio-quiet quasars including 207 radio-quiet
quasars with measured radio luminosity. Then we visually check
these spectra one by one.

At last, we obtain a sample of 3772 quasars with better
multi-components model of \hb and \oiii lines, including 3466
radio-quiet quasars (hereafter "RQ total sample"), 306 radio-loud
quasars (hereafter "RL sample"). Most objects in these 3466
radio-quiet quasars only have upper-limits of the radio luminosity
and the radio loudness, 181 radio-quiet quasars (hereafter "RQ
sample") have the measurements of the radio luminosity and the
radio loudness. We use the radio-quiet sample as the control
sample to discuss the $\mbh-\sigma_*$ in radio-loud quasars.

\section{SMBH Mass, Eddington ratio and stellar velocity dispersion}
The BLRs size is calculated from the monochromatic luminosity at
5100\AA\ (\lv) or the \hb luminosity by the following formulae
(Kaspi et al. 2005):
\begin{eqnarray}
R^{\rm \lv}_{\rm BLR}=(22.3\pm 2.1)\left(\frac{\lv}{10^{44}~
\ergs}\right)^{0.69\pm0.05} \rm lt-days
\end{eqnarray}

\begin{eqnarray}
R^{\rm L_{\hb}}_{\rm BLR}=(82.3\pm 7.0)\left(\frac{L_{\rm
\hb}}{10^{43}~ \ergs}\right)^{0.80\pm0.11} \rm lt-days
\end{eqnarray}

We use the FWHM of the broad \hb line ($\rm FWHM_{\hb}$) to trace
the BLRs virial velocity $v_{\rm BLR}=\sqrt{f}\times \rm
FWHM_{\hb}$, $f$ is the calibration factor. If BLRs cloud is
disk-like with a inclination of $\theta$ (Wills \& Browne 1986),
\begin{equation}
{\rm FWHM_{\rm H \beta}}=2(v_{\rm r}^{2}+v_{\rm BLR}^{2} \rm
sin^{2}\theta)^{1/2}
\end{equation}
where $v_{\rm r}$ is the random isotropic component. We can then
calculate the SMBH masses by $M_{\rm BH} = \frac{R_{\rm BLR}v_{\rm
BLR}^2}{G}$ (Kaspi et al. 2000; Kaspi et al. 2005):
\begin{eqnarray}
M_{\rm BH} =f\times 4.35 \times 10^6 \left(\frac{\rm     \nonumber
FWHM_{\hb}}{10^{3} \kms}\right)^2\left(\frac{\lv}{10^{44}\ergs}
\right)^{0.69}  \msun.                                 \\
\end{eqnarray}

\begin{eqnarray}
M_{\rm BH}=f\times 1.61 \times 10^7 \left(\frac{\rm  \nonumber
FWHM_{\hb}}{10^{3}\kms}\right)^2\left(\frac{L_{\rm
\hb}}{10^{43}\ergs}
\right)^{0.80}  \msun.                                 \\
\end{eqnarray}
If assuming $v_{\rm r} \ll v_{\rm BLR}$, and the random orbits of
BLRs clouds, $f=0.75$. Onken et al. (2004) did a calibration by
the $\mbh - \sigma_*$ relation and suggested $f\approx1.4$ (see
also Collin et al. 2006; Dasyra et al. 2007). In our mass
calculation, we adopt the random orbits of BLRs clouds and
$f=0.75$.

We calculate the Eddington ratio, i.e., the ratio of the
bolometric luminosity ($L_{\rm bol}$) to the Eddington luminosity
($L_{\rm Edd}$), where $L_{\rm Edd}=1.26 \times 10^{38} (M_{\rm
BH}/ \msun) \ergs$. The bolometric luminosity is calculated from
the monochromatic luminosity at 5100\AA\ , $L_{\rm bol}=c_{\rm
B}\lv$, where we adopt the correction factor $c_{\rm B}$ of 9
(Kaspi et al. 2000; Marconi et al. 2004; Richards et al. 2006;
Netzer \& Trakhtenbrot 2007).

We use the gas velocity dispersion of the narrow/core \oiii
component from NLRs to trace the host stellar velocity dispersion,
$\sigma^{n}_{\rm[O III]}=\sqrt{\sigma_{\rm obs}^2-[\sigma_{\rm
inst}/(1+z)]^2}$, where $\sigma_{\rm obs}={\rm FWHM}^{n}_{\rm [O
III]}/2.35$, $z$ is the redshift (Bian et al. 2006). For SDSS
spectra, the mean value of instrument resolution $\sigma_{\rm
inst}$ is 60 \kms\ for \oiii (e.g. Greene \& Ho 2005a).

\begin{figure}
\begin{center}
\includegraphics[width=11cm]{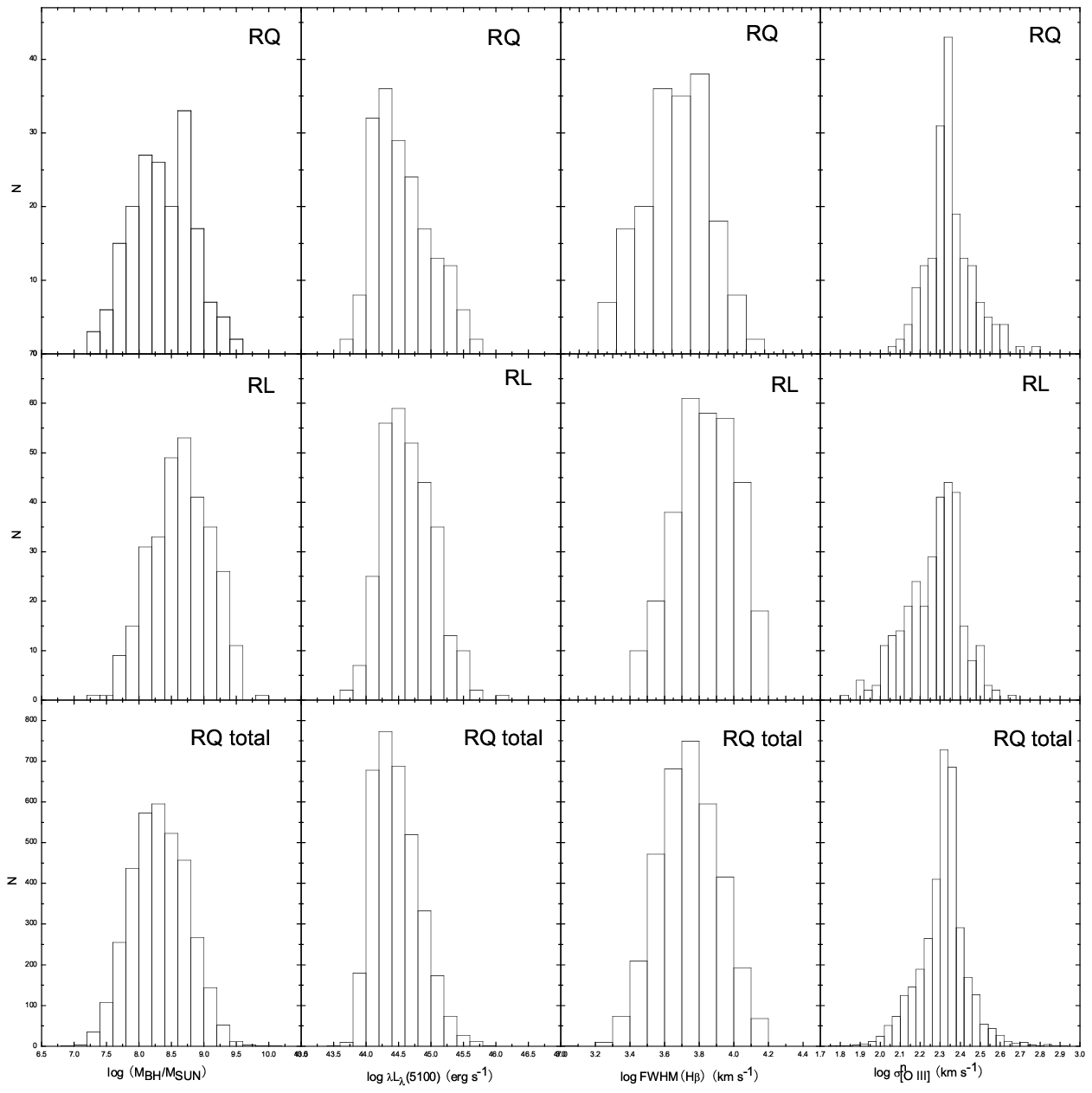}
\caption{The distributions of $\mbh$, $\lv$, $\rm FWHM_{\rm
H\beta}$, $\rm \sigma^{n}_{\rm [O III]}$ for 181 radio-quiet
quasars with radio loudness (top), 306 radio-loud quasars with
measured radio luminosity (middle), total 3466 radio-quiet quasars
(bottom).}
\end{center}
\end{figure}

In Figure 2, we present the distributions of the SMBH mass, $\lv$,
$\rm FWHM_{\rm H\beta}$, $\rm \sigma^{n}_{\rm [O III]}$ for 181
radio-quiet quasars with radio loudness (top), 306 radio-loud
quasars with measured radio luminosity (middle), total 3466
radio-quiet quasars (bottom). The mean of SMBH mass is $8.65\pm
0.03$ with a standard deviation of $0.45$ for RL sample of 306
radio-loud quasars, $8.36\pm 0.04$ with a standard deviation of
$0.48$ for RQ sample of 181 radio-quiet quasars with reliable
radio luminosity, $8.32\pm 0.01$ with a standard deviation of
$0.43$ for total 3466 radio-quiet quasars. Radio-loud quasars have
larger SMBH masses, and there is only a few objects with mass less
than $10^8 \msun$ (see Figure 3), which is consistent with the
results of McLure \& Jarvis (2004). Radio-loud quasars have
smaller Eddington ratios, respect to radio-quiet quasars (see
Table 1). We find that, for radio-loud quasars, the mean of \hb
FWHM is $7493\pm165$ \kms with a standard deviation of $2882$
\kms, the mean of \llog \lv is $44.86\pm0.03$ \ergs with a
standard deviation of $0.45$; for radio-quiet quasars, the mean of
\hb FWHM is $5780\pm176$ \kms with a standard deviation of $2389$
\kms, the mean of \llog \lv is $44.81\pm0.01$ \ergs with a
standard deviation of $0.46$. Radio-loud quasars tend to have
larger \hb FWHM and \lv, leading larger SMBH masses (Sulentic et
al. 2000).

\section{$\mbh-\sigma_*$ relation}
\subsection{The mass deviation from $\mbh- \sigma_*$ relation}

\begin{figure}
\begin{center}
\includegraphics[width=9cm]{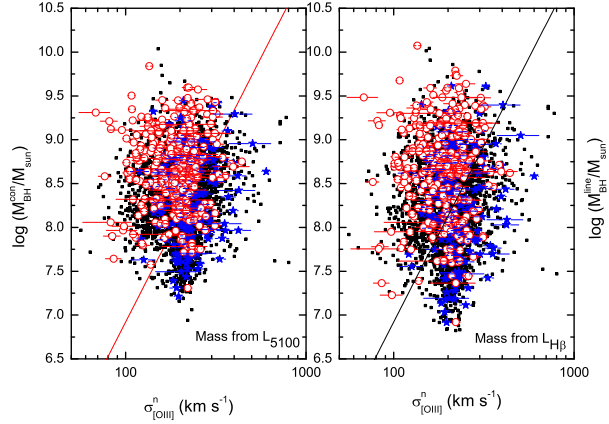}
\caption{The $\mbh- \sigma_*$ relation for radio-loud and
radio-quiet quasars. Red circle denotes radio-loud quasars, blue
star denotes radio-quiet quasars with measured radio luminosity,
black square denotes the radio-quiet quasars with upper-limit of
the radio luminosity. The mass in the left panel is derived from
\lv, and the mass in the right panel is derived form \hb
luminosity.}
\end{center}
\end{figure}

\begin{figure}
\begin{center}
\includegraphics[width=9cm]{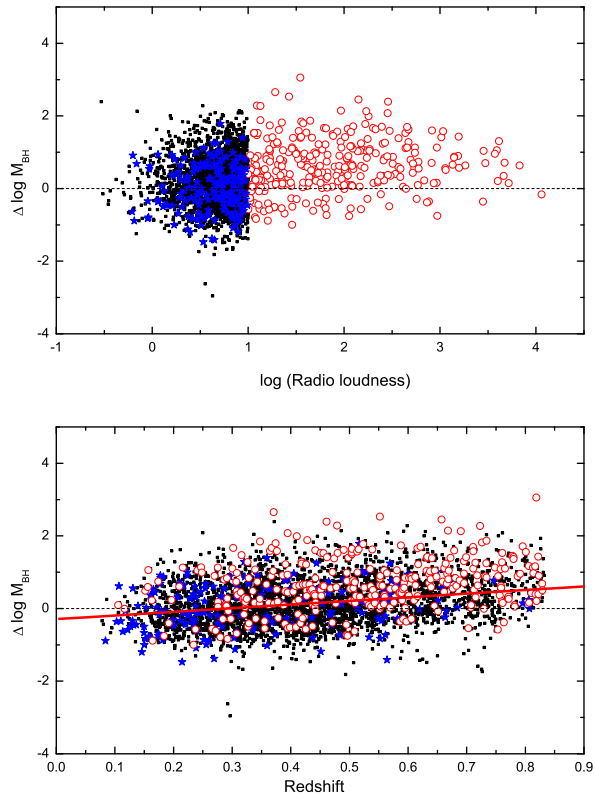}
\caption{Top: The deviation of the SMBH mass from the Tremaine's
$\mbh- \sigma_*$ relation in fig 2. versus the radio loudness. The
dash line denotes $\Delta \llog \mbh=0$. Bottom: The deviation of
the SMBH mass from the Tremaine's $\mbh- \sigma_*$ relation in fig
2. versus the redshift. The red solid line denotes our best fit
for all radio-quiet quasars. Symbols as Figure 3. }
\end{center}
\end{figure}

\begin{figure}
\begin{center}
\includegraphics[width=9cm]{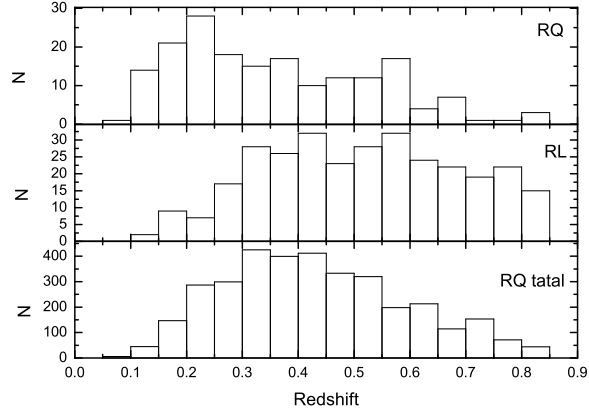}
\caption{The redshift distributions for 181 radio-quiet quasars
with radio loudness (top), 306 radio-loud quasars with measured
radio luminosity (middle), total 3466 radio-quiet quasars
(bottom).}
\end{center}
\end{figure}

In Figure 3, we show $\mbh- \sigma_*$ relation for radio-loud and
radio-quiet quasars. The solid line in Figure 3 is the $\mbh-
\sigma_*$ relation in normal nearby galaxies given by Tremaine et
al. (2002), $\mbh (\sigma_*) = 10^{8.13}[\sigma_{*}/(200 \ \rm km
s^{-1})]^{4.02} ~~\msun $. In the left and right panels of Figure
3, \lv and $L_{\rm \hb}$ are used to calculate the SMBH mass,
respectively. In Figure 3, the correlation between \mbh and
$\sigma^{n}_{\rm [O III]}$ is very weak for larger SDSS quasars
sample. It is possibly due to the accuracy of the stellar velocity
dispersion derived from the narrow/core \oiii line-width. However,
it is obvious that the sample of radio-loud quasars deviated much
from the solid line respect to that for the sample of radio-quit
quasars. It is consistent with our previous result (Bian \& Zhao
2004).

We calculate the black hole mass deviation $\Delta \llog \mbh$
from the solid line defined by Tremaine et al. (2002), $\Delta
\llog \mbh = \llog \mbh(\rm \hb) -\llog \mbh(\sigma_*)$, where
$\sigma_*$ is adopted to be $\sigma^{n}_{\rm [O III]}$. For the
mass derived from \lv, the mean of $\Delta \llog \mbh1$ is
$0.65\pm 0.04$ with a standard deviation of $0.71$ for RL sample
of 306 radio-loud quasars, $0.04\pm 0.04$ with a standard
deviation of $0.63$ for RQ sample of 181 radio-quiet quasars with
reliable radio luminosity, $0.14\pm 0.01$ with a standard
deviation of $0.62$ for total 3466 radio-quiet quasars. We find
that they are almost the same for the case of the mass derived
from $L_{\rm \hb}$. In the next analysis, we just consider the
mass and Eddington ratio calculated from \lv.

In the top panel of Figure 4, we plot the deviation of the SMBH
mass from  $\mbh- \sigma_*$ relation versus the radio loudness. It
is obvious that the deviation tends to be larger when radio
loudness becomes larger. In the bottom panel of Figure 4, we also
plot the deviation of the SMBH mass from the $\mbh- \sigma_*$
relation versus the redshift. We find a weak correlation between
the mass deviation and the redshift for radio-quiet quasars. The
simple least-square regression gives: $\Delta \llog \mbh =
(1.00\pm 0.06)z-(0.29\pm 0.03)$. The correlation coefficient $R$
is 0.26, with a probability of $p_{\rm null} < 10^{-4}$ for
rejecting the null hypothesis of no correlation. In Figure 5, we
show the redshift distributions for radio-quiet and radio-loud
quasars. The radio-loud quasars (red circles) have larger
redshifts relative to the radio-quiet quasars (blue stars) (see
Figure 5).

In Table 1, we show the mean values of the masses and the
Eddington ratios in different redshift bins for different samples.

\subsection{Uncertainties}

There are some factors to account for the uncertainty of the SMBH
mass calculation: the uncertainties of \hb, \oiii line widthes,
\lv, $L_{\rm \hb}$ when the multi-components are used to model
SDSS spectra; the system errors in equations (1-5) from the
uncertainties of the BLRs geometry and dynamics. The uncertainty
of our calculated SMBH mass is about 0.5 dex. The uncertainty of
the Eddington ratio is about 0.5 dex or more. For radio-loud
quasars, we should account two effect: the relativistic beaming
effect on the optical continuum and the orientation of BLRs. The
total \hb luminosity instead of \lv is used to account for the
first effect. We find the effect is small in our sample, and there
is no correlation between the H$\beta$ EW and the radio loudness
(e.g. Wu et al. 2004). Lacy et al. (2001) made a small correction
of the orientation of BLRs by a factor of $R^{0.1}_{\rm c}$, where
$R_{\rm c}$ is the ratio of core to extended radio luminosity.
They adopted $R_{\rm c}=0.1$ for steep-spectrum quasars and
$R_{\rm c}=10$ for flat-spectrum quasars if $R_{\rm c}$ is not
measured. It will lead to the uncertainties of $\Delta \llog \mbh$
about 0.2 dex.

The fibers in the SDSS spectroscopic survey have a diameter of 3"
on the sky. The SDSS spectra of lower-redshift quasars possibly
have obvious stellar light contribution, which can be used to
directly measure the stellar velocity dispersion (e.g., Kauffmann
et al. 2003; Heckman et al. 2004; Bian et al. 2006). For luminous
quasars ($M_{i}<-22$), the stellar light contribution can be
omitted or has little effect on the mass calculation (e.g., Vanden
Berk et al. 2006).

It is possible that jet can have a dynamical effect on the NLRs
and may have a systematically different effect on the \oiii
profile (Nelson \& Whittle 1996). However, considering the \oiii
profile broadening by jet, the correction of \oiii gas velocity
dispersion will lead the radio-loud quasars to deviate much more
from the $\mbh- \sigma_*$ relation.

We select quasars with the EW of narrow \oiii component larger
than 1.5, EW errors of \hb and \oiii$\lambda \lambda 4959, 5007$
less than $100\%$. Different criteria would lead to different
number of quasars (such as error of EWs less than 5\%, 50\%, or
100\%, $\chi^2 <4$). However, we find that the main results don't
change. In the next subsection, we will discuss the luminosity
bias in detail.

\subsection{The mass deviation from the luminosity bias}

\begin{figure}
\begin{center}
\includegraphics[width=9cm]{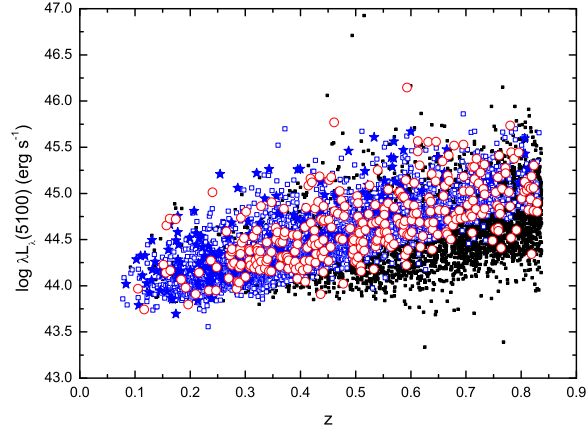}
\caption{\lv versus $z$. Open circles denote RL sample; blue stars
denote RQ QSOs with measured radio loudness; blue squares denote
RQ QSOs with upper-limits of radio loudness; small black squares
denote all 9753 SDSS DR3 QSOs with $z < 0.83$. It is obvious that
some faint objects are missed in our selection.}
\end{center}
\end{figure}

\begin{tiny}
\begin{table*}
\begin{center}
\begin{tabular}{lcccccccccccc} \hline \hline
z    &     N   &  log(\lv \ergs)  & log (\mbh/\msun) & log (\lb/\ledd) & $\Delta $log \mbh &   ${\rm log} L_{\rm cut}/L_0$   & $\Delta$log\mbh$^{simu}$\\
(1)     &   (2)      &  (3)   &  (4)        &   (5)              &  (6)&(7)& (8)  \\
\hline
\textbf{RQ Total} &      &   &   &  &  &                &     \\
0.1-0.2& 198  &  $44.08\pm 0.17$ &  $7.81\pm 0.41$ & $-1.05\pm 0.40$ & $-0.08\pm 0.54$ &  -1.38   & 0.04\\
0.2-0.3  & 586  &  $44.17\pm 0.21$ &  $7.92\pm 0.39$ & $-1.06\pm 0.37$ & $-0.02\pm 0.61$ &  -1.30   & 0.06\\
0.3-0.4  & 824  &  $44.32\pm 0.24$ &  $8.05\pm 0.43$ & $-1.02\pm 0.36$ & $ 0.07\pm 0.58$ &  -1.18   & 0.09\\
0.4-0.5  & 745  &  $44.48\pm 0.22$ &  $8.24\pm 0.41$ & $-0.98\pm 0.36$ & $ 0.08\pm 0.61$ &  -0.98   & 0.15\\
0.5-0.6  & 518  &  $44.67\pm 0.24$ &  $8.49\pm 0.43$ & $-0.97\pm 0.37$ & $ 0.30\pm 0.61$ &  -0.74   & 0.21\\
0.6-0.7  & 327  &  $44.82\pm 0.23$ &  $8.62\pm 0.43$ & $-0.93\pm 0.33$ & $ 0.37\pm 0.60$ &  -0.58   & 0.24\\
0.7-0.83 & 267  &  $45.04\pm 0.23$ &  $8.80\pm 0.39$ & $-0.87\pm 0.34$ & $ 0.41\pm 0.63$ &  -0.38   & 0.26\\
\textbf{RQ}  &      &   &   &  &  &                &     \\
0.1-0.4& 114  &  $44.35\pm 0.32$ &  $8.21\pm 0.32$ & $-1.01\pm 0.39$ & $-0.04\pm 0.63$ &  -1.14   & 0.10\\
0.4-0.82& 67   &  $44.94\pm 0.34$ &  $8.62\pm 0.41$ & $-0.83\pm 0.32$ & $ 0.17\pm 0.61$ &  -0.50   & 0.25\\
\textbf{RL } &      &   &   &  &  &                &     \\
0.1-0.3  & 35   &  $44.25\pm 0.29$ &  $8.25\pm 0.38$ & $-1.15\pm 0.42$ & $ 0.01\pm 0.68$ &  -1.22   & 0.08\\
0.3-0.5  & 109  &  $44.49\pm 0.31$ &  $8.51\pm 0.45$ & $-1.17\pm 0.39$ & $ 0.52\pm 0.80$ &  -0.98   & 0.15\\
0.5-0.7  & 106  &  $44.78\pm 0.35$ &  $8.78\pm 0.38$ & $-1.14\pm 0.32$ & $ 0.80\pm 0.63$ &  -0.66   & 0.23\\
0.7-0.83 & 56   &  $44.98\pm 0.29$ &  $8.91\pm 0.39$ & $-1.08\pm
0.33$ & $ 0.95\pm 0.75$ &  -0.46   & 0.26\\
\hline
\end{tabular}
\caption{The mean quantities in different redshift bins for
different samples. $L_{\rm cut}$ is calculated from the QSOs
luminosity function (Boyle et al. 2000) to make the mean
luminosity of the kept QSOs ($L >L_{\rm cut}$ ) equal to the
observed mean luminosity in different redshift bins.
$L_{0}=0.3L_{\rm Edd}(M^*_{\rm gal})$, where $M^*_{\rm
gal}=10^{11}\msun$ in the galaxy mass function $\Phi(M_{\rm
gal})=\Phi^*(M_{\rm gal}/M^*_{\rm gal})^{-a}  e^{-M_{\rm
gal}/M_{\rm gal}^*}$ (Drory et al. 2005). }
\end{center}
\end{table*}
\end{tiny}

Salviander et al. (2007) also used SDSS DR3 quasars to discuss the
cosmological evolution of $\mbh- \sigma_*$ relation.  After
carefully consider the selection biases and intrinsic scatter in
the $\mbh- \sigma_*$ relation, they suggested that $\mbh-
\sigma_*$ relation appears to evolve with redshift. Netzer \&
Trakhtenbrot (2007) also found the nonlinear $\mbh- \sigma_*$
relation with the different slopes for different redshift bins.
During our process of selecting objects, the line fitting favor
the brighter objects (i.e. luminosity bias, see Figure 6).
Following the work of Salviander et al. (2007), we calculate the
contribution of $\Delta \llog \mbh$ from this luminosity bias. We
calculate the mean observed luminosity in different redshift bins
for our different samples (i.e. RQ total sample; RQ sample; RL
sample). Using the QSOs luminosity function (Boyle et al., 2000),
we calculate the cut luminosity to make the mean luminosity of the
kept QSOs ($L
>L_{\rm cut}$ ) equal to the observed mean luminosity in different
redshift bin. Then we do the simulation to calculate the
contribution of $\Delta \llog \mbh$ from this luminosity bias (for
detail in Salviander et al. 2007). We obtained a formulae: $\Delta
\llog \mbh^{\rm simu} = 0.292+0.1138x+0.265x^2+0.480x^3+0.182x^4$,
where $x$ is $\llog (L_{\rm cut}/L_0)$, $L_0=0.3 \ledd (M^*_{\rm
gal})$, $M^*_{\rm gal}=10^{11} \msun$ (Drory et al. 2005). We find
that the mass deviation form the luminosity bias is monotonously
increased with the redshift (see Col.(8) in Table 2). Table 1
shows our results. Col.(1) is the redhsift bin; Col.(2) is the
number in the redshift bin; Col.(3)-(5) are the mean values of
5100\AA luminosity, mass, Eddington ratio; Col.(6) is the mean
mass deviation from $\mbh - \sigma_*$ relation; Col.(7) is $\llog
(L_{\rm cut}/L_0)$; Col.(8) is our simulated mass deviation for
different cut luminosity in different redshift bins.

For RQ total sample, the luminosity bias can interpret most amount
of $\Delta \llog \mbh$. For the highest redshift bin of
$0.7<z<0.83$, $\Delta \llog \mbh$ is about 0.15 dex after
correction the effect of luminosity bias. This 0.15 dex is
possibly the $\mbh-\sigma_*$ cosmological evolution in this
highest redshift bin, which is very consistent with the result of
Salviander et al. (2007). However, we should note that the
standard deviation of $\Delta \llog \mbh$ in different bins is
about 0.6dex, which is very larger than this 0.15dex. For RQ
sample, the observed $\Delta \llog \mbh$ can be completed
contributed from the luminosity bias, which is possibly due to the
smaller numbers of this sample. Therefore, we think there is no
obvious deviation from $\mbh- \sigma_*$ relation considering the
\mbh uncertainties and the luminosity bias.

For RL sample, after corrected the contribution from the
luminosity bias, $\Delta \log \mbh$ is still large (about 0.69 dex
in $0.7<z<0.83$) and there is a trend that $\Delta \llog \mbh$
becomes larger for larger redshift bin. Considering the possible
$\mbh-\sigma_*$ cosmological evolution (0.15 dex in $0.7<z<0.83$),
for radio loud QSOs, there are still 0.54 dex deviation in
$0.7<z<0.83$. Bonning et al. (2005) suggested that narrower \oiii
for radio loud quasars is responsible for this deviation from the
$\mbh-\sigma_*$ relationship, and it is not the effect involving
\mbh. The cause of this deviation is unclear.

\section{Origin of radio luminosity}

\subsection{$L_{\rm 5 GHz} - L_{\rm [O \sc{III}]}$ relation}
\begin{figure}
\begin{center}
\includegraphics[width=10cm]{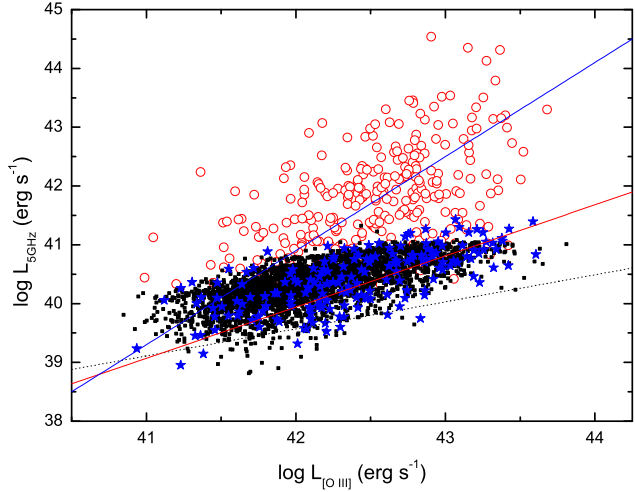}
\caption{The radio luminosity versus the \oiii luminosity. Red
circle denotes radio-loud quasars, blue star denotes radio-quiet
quasars with measured radio luminosity, black square denotes the
radio-quiet quasars with upper-limit of the radio luminosity. The
red solid line is the BCES bisector result for radio-quiet quasars
(blue stars). The blue solid line is the BCES bisector result for
radio-loud quasars (red circles). The dash line is the best fit of
radio-quiet AGNs found by Xu et al. (1999).}
\end{center}
\end{figure}
The relation between the radio luminosity and the optical/X-ray
luminosity, which provides the connection between the jet and
accretion power, have been discussed by many group (e.g. Xu et al.
1999; Ho 2002; Wang et al. 2004; Panessa et al. 2007; Sikora
2007). In Figure 7, we show the radio luminosity at 5GHz versus
the total \oiii luminosity. These two luminosities are all related
to the redshift. By the partial Kendall's $\tau$ correlation test,
we do a partial correlation analysis with redshift as the test
variable(Akritas \& Siebert 1996). For the RQ sample of 181
radio-quiet quasars, partial Kendall's $\tau$ correlation is
0.237, variance is 0.0304, and the probability of null hypothesis
is $6.3\times 10^{-15}$ . For RL sample of 306 radio-loud quasars,
$\tau$ correlation is 0.251, variance is 0.0456, and the
probability of null hypothesis is $3.7\times 10^{-8}$. We use the
bivariate correlated errors and intrinsic scatter (BCES)
regression method {\footnote {This is not a symmetric regression
used by Merloni et al. 2003. For detail in section 5.2}} of
Akritas \& Bershady (1996) (see also Isobe et al. 1990) to find
the relation between $L_{\rm [O \sc ~III]}$ and $L_{\rm 5 GHz}$,
and adopt the BCES bisector result (e.g. Kaspi et al. 2005). For
RQ sample of 181 radio-quiet quasars with measured radio
luminosity, the BCES bisector result: $\llog L_{\rm 5 GHz} =
(0.87\pm 0.04) \llog L_{\rm [\rm O {\sc ~III}]} + (3.40\pm 1.87)$
(red dash line in Figure 7). For RL sample of 306 radio-loud
quasars, $\llog L_{\rm 5 GHz} = (1.60\pm 0.08) \llog L_{\rm [\rm O
{\sc ~III}]} - (26.30\pm 3.39)$.

In Figure 7, considering the errors of the intercept, our best
fits for radio-quiet quasars is consistent with the result found
by Xu et al. (1999) (also see Ho \& Peng 2001) : $\llog L_{\rm 5
GHz} = (0.45\pm 0.07) \llog L_{\rm [\rm O {\sc ~III}]} + (20.25\pm
0.6)$ (black dot line in Figure 7). In the plot of radio
luminosity versus the optical/X-ray nuclear luminosity, the
separation of radio-loud and radio-quiet quasars from SDSS DR3 is
not too clear as other's results (Xu et al. 1999; Terashima \&
Wilson 2003; Sikora et al. 2007). The difference is possibly due
to the selection effect by different wavelength bands.

\oiii luminosity is usually assumed to be proportional to the
accretion rate and this correlation can be explained in a model of
accelerated and collimated jet by magnetic field (Xu et al. 1999).
Apart from the dependence on the accretion rate, the radio
luminosity possibly dependents on the central SMBH properties,
mass or spin (e.g., Sokira et al. 2007 and refs. therein), which
we will discuss in the next section.

If we use the tight correlation between X-ray luminosity and \oiii
luminosity (e.g., Xu et al. 1999), $\llog L_{\rm x}=1.01\llog
L_{[\rm O III]}+1.6$, the relation between $L_{\rm 5 GHz}$ and
$L_{\rm [O \sc{III}]}$ can be transformed to the relation between
$L_{\rm 5 GHz}$ and $L_{\rm x}$: $L_{\rm 5 GHz} \propto L_{\rm
x}^{0.86\pm 0.06}$ for RQ sample and $L_{\rm 5 GHz} \propto L_{\rm
x}^{1.58\pm0.10}$ for RL sample. There exists obvious different on
the slope for radio-quiet and radio-loud quasars. For low
luminosity AGNs, Panessa et al. (2007) suggested a correlation,
$L_{\rm x} \propto L_{\rm 5GHz}^{0.97}$, their index is between
ours for radio-quiet and radio-loud quasars. If we use the
correlation suggested by Netzer et al. (2006), $L_{\rm O III}
\propto L_{\rm x}^{0.704\pm 0.06}$, the relation we found between
$L_{\rm 5 GHz}$ and $L_{\rm [O \sc{III}]}$ can be transformed to
$L_{\rm 5 GHz} \propto L_{\rm x}^{0.61\pm 0.04}$ for RQ sample and
$L_{\rm 5 GHz} \propto L_{\rm x}^{1.11\pm0.07}$ for RL sample. The
radio luminosity is often assumed coming from the relativistic
electrons powered by a jet. The result of RL sample is consistent
with that of Panessa et al. (2007). The X-ray emission is often
assumed coming from both the accretion flow and the relativistic
jet, dominated by accretion flow at high accretion rate, and
dominated by jet emission at low accretion rate. (Gallo et al.
2003; Yuan \& Cui 2005). This relation between $L_{\rm 5 GHz}$ and
$L_{\rm x}$ in different accretion rates can be explained in the
jet-dominant X-ray models (e.g., Fender et al. 2003; Gallo et al.
2003; Heinz 2004; Yuan \& Cui 2005).

\subsection{The radio luminosity dependence on the SMBH mass and the Eddington ratio}
\begin{figure}
\begin{center}
\includegraphics[width=10cm]{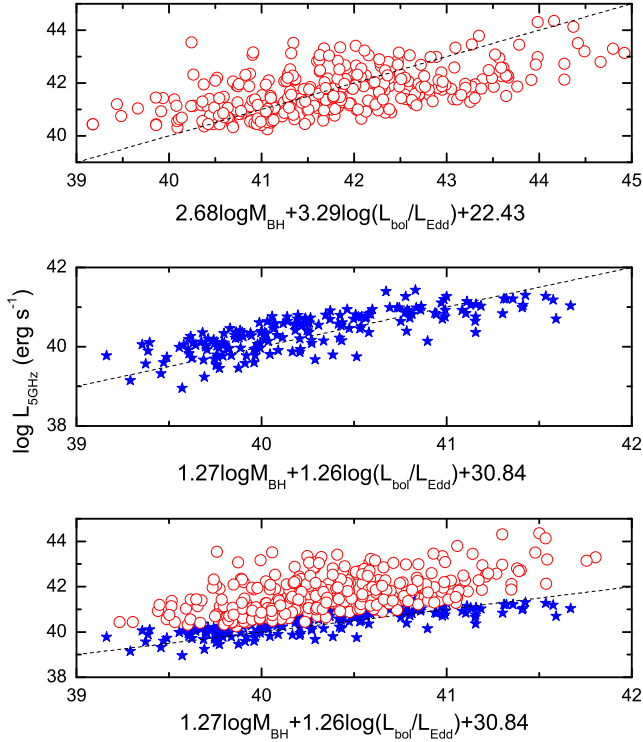}
\caption{The radio luminosity dependence on the SMBH mass and the
Eddington ratio. The indexes are adopted from the mean values in
brackets in Table 1. Top panel is for radio-loud quasars, middle
panel is for radio-quiet quasars with measured radio luminosity,
bottom panel is for them all. Symbols as Figure 7. The dash lines
denote 1:1.}
\end{center}
\end{figure}

\begin{tiny}
\begin{table*}
\begin{center}
\caption{The mean quantities in different redshift bins for
different samples. $a,b_1,b_2$ is defined by: $\llog L_{\rm 5
GHz}=a+ b_1\llog M_{\rm BH} +b_2 \llog(\lb/\ledd)$. For different
samples, the first line is for the result by $\chi^2$
minimization, and in the second line, quantities in brackets are
the mean values of $a,b_1,b_2$ by ASURV, the last three lines are
results considering different dependent variable by ASURV.}
\begin{tabular}{lcccccccccccc} \hline \hline
Dependent variable &   a &   $b_1$ &  $b_2$  & R-sqaure \\
(1) & (2)&(3)&(4)&(5)\\
\hline \textbf{RQ } & $30.9^{+1.20}_{-1.80}$ &  $1.28^{+0.23}_{-0.16}$ & $1.29^{+0.31}_{-0.24}$\\
&  ($30.84\pm 2.21$) &  ($1.27\pm 0.29$) &  ($1.26\pm 0.53$)       &   &      \\
$L_{\rm 5GHz}$ &  $33.18\pm 0.43$ &  $0.95\pm 0.05$ &  $0.81\pm 0.07$ &  0.64\\
    \mbh      &  $28.79\pm 1.54$ &  $1.52\pm 0.02$ &  $1.12\pm 0.08$ &  0.72\\
    \lb/\ledd  &  $30.54\pm 3.04$ &  $1.35\pm 0.09$ &  $1.85\pm 0.02$ &  0.57\\
\hline \textbf{RL}& $19.7^{+5.40}_{-3.90}$ &  $3.10^{+0.60}_{-0.70}$ & $4.18^{+1.40}_{-1.10}$\\
&  ($22.43\pm 10.4$)&  ($2.68\pm 1.30$) &  ($3.29\pm 1.94$)       &   &      \\
$L_{\rm 5GHz}$ &  $32.62\pm 0.82$ &  $1.24\pm 0.10$ &  $1.38\pm 0.13$ &  0.35\\
    \mbh      &  $11.77\pm 3.54$ &  $3.85\pm 0.40$ & $3.23\pm 0.19$  &  0.53\\
    \lb/\ledd  &  $22.89\pm 3.79$ &  $2.95\pm 0.16$ &  $5.26\pm 0.05$ &  0.49\\
\hline \textbf{RL+RQ}& $10.0^{+8.70}_{-4.20}$ &  $4.30^{+0.70}_{-0.80}$ & $5.15^{+2.32}_{-1.69}$\\
&  ($15.65\pm 12.93$)& ($3.46\pm 1.81$) &  ($4.11\pm 3.81$)         &   &      \\
$L_{\rm 5GHz}$ &  $30.45\pm 0.73$ &  $1.38\pm 0.09$ &  $0.92\pm 0.12$ &  0.31\\
    \mbh      &  $ 6.57\pm 2.74$ &  $4.34\pm 0.04$ &  $3.09\pm 0.17$ &  0.53\\
    \lb/\ledd  &  $ 9.92\pm 4.67$ &  $4.67\pm 0.25$ &  $8.33\pm 0.08$ &  0.40\\
\hline
\end{tabular}
\end{center}
\end{table*}
\end{tiny}

It is suggested that the radio luminosity/radio loudness is relate
to the SMBH masses (e.g.,  Laor 2000). We calculate the radio
luminosity dependence on the SMBH mass and the Eddington ratio,
i.e. $\llog L_{\rm 5 GHz}=a+ b_1\llog M_{\rm BH} +b_2
\llog(\lb/\ledd)$ (see Figure 8).

We firstly do the multiple regression with ASURV Rev 1.2
(LaValley, Isobe \& Feigelson 1992 and refs. therein) for RQ
sample, RL sample and RL+RQ sample. In order to avoid the
non-symmetric regression, $a,b_1,b_2$ are adopted the mean values
when we use different variable as the dependent variables in the
multiple regressions (see Col.(1) in Table 1). In all the multiple
regressions, the probability  for rejecting the null hypothesis of
no correlation is $p_{\rm null} < 10^{-4}$. The R-Square
correlation coefficient for RQ sample is larger than other two
samples (see Table 2, Figure 8).


We also do the symmetric multivariate regression analysis, through
the equation $y=a+b_1 x_1 + b_2 x_2$, directly by the $\chi^2$
estimator, $\chi^2=\sum_i\frac{(y_{i}-a-b_1x_{1i}+b_{2}
x_{2i})^2}{\sigma_{yi}^2+(b_1\sigma_{x_{1i}})^2+(b_2\sigma_{x_{2i}})^2}$,
(Press et al. 1992; Tremaine et al., 2002; Merloni et al., 2003),
where $\sigma$ are the corresponding uncertainties. Considering
the same uncertainties ($\sigma$) of radio luminosity, mass, and
the Eddington ratio (Tremaine et al., 2002; Merloni et al., 2003),
we re-normalized these uncertainties to make the minimum
$\chi^2/n_{\rm dof}$ of unity, the results are listed in first
lines for different sample in Table 1.

Considering the errors of $a,b_1,b_2$ in Tables 2, the results
from ASURV and $\chi^2$ are consistent very well. Therefore, in
the next analysis, we adopt the values of $a,b_1,b_2$ from
$\chi^2$ estimator (Table 1), i.e. $L_{\rm 5 GHz} \propto
\mbh^{1.28^{+0.23}_{-0.16}} (\lb/\ledd)^{1.29^{+0.31}_{-0.24}}$
for RQ sample, and $L_{\rm 5 GHz} \propto
\mbh^{3.10^{+0.60}_{-0.70}} (\lb/\ledd)^{4.18^{+1.40}_{-1.10}}$for
RL sample, and $L_{\rm 5 GHz} \propto \mbh^{4.30^{+0.70}_{-0.80}}
(\lb/\ledd)^{5.15^{+2.32}_{-1.69}}$ for RL+RQ sample.

Ho (2002) suggested a correlation between the nuclei radio
loudness and the Eddington ratio (Gallo et al. 2003; Greene et al.
2006; Sikora et al. 2007; Panessa et al. 2007). We also use the
multiple regression by ASURV to search the radio loudness
dependence on the SMBH mass and the Eddington ratio (set radio
luminosity as the dependent variable). However, the R-Square
correlation coefficient is very low for RL sample and RQ sample.
For the RL+RQ sample, we find a weak correlation between the radio
loudness and the SMBH mass (the simple least-square correlation
R=0.26), much weaker correlation between the radio loudness and
the Eddington ratio (R=-0.15). The range of Eddington ratio is
between 0.01 to 1 for our RL sample and RQ sample. And our sample
is composed by broad line type I quasars, which just fill the gaps
between two sequences in plot of radio loudness versus the
Eddington ratio (see Figure 3 in Sikora et al. 2007). When we
research the disk-jet connection model, X-ray luminosity is maybe
a better tracer of SMBHs accretion power than the optical
luminosity (e.g. Panessa et al. 2007). We also should pay more
attention on narrow-line Seyfert 1 galaxies with larger Eddington
ratios in this kind of plot (e.g. Zhou \& Wang 2002; Whalen et al.
2006; Komossa et al. 2006).

For scale-free jet physics, Heinz \& Sunyaev (2003) derived the
dependence of the accretion-powered jets flux ($f_{v}$) upon the
SMBH  mass and the dimensionless accretion rates for different
accretion scenarios (see their Table 1). For
radiation-pressure-supported standard accretion disk, $f_{v}
\propto \mbh^{17/12-\alpha/3}$; for gas-pressure-supported
standard accretion disk, $f_{v} \propto \mbh^{(187-32\alpha)/120}
\mdot^{(17/12+2\alpha/3)4/5}$; for ADAF, $f_{v} \propto
\mbh^{17/12-\alpha/3} \mdot^{17/12+2\alpha/3}$, where $\alpha$ is
the radio spectral index. Assuming $\alpha=0.5$, for
radiation-pressure-supported standard accretion disk, $f_{v}
\propto \mbh^{1.25}$; for gas-pressure-supported standard
accretion disk, $f_{v} \propto \mbh^{1.43} \mdot^{1.40}$; for
ADAF, $f_{v} \propto \mbh^{1.25} \mdot^{1.75}$. Considering large
scatter in  $b_1,b_2$, our results are consistent with above the
radio origin of scale-free jet model. However, by our data, we
can't distinguish the different disks for radio-quiet and
radio-loud quasars in this accretion-powered jet model.

\section{conclusions}
With the large number of quasars in SDSS DR3 catalog, we use the
multi-components to model the SDSS spectra and calculate the SMBH
masses. Combined with the radio properties from FIRST, we obtained
a sample of 3772 quasars with reliable SMBH masses, including 306
radio-loud quasars, 3466 radio-quiet quasars with measured radio
luminosity or upper-limit of radio luminosity (181 radio-quiet
quasars with measured radio luminosity). Two main results are
suggested: (1) The radio-loud quasars deviate much from the $\mbh-
\sigma_*$ relation of the nearby normal galaxies defined by
Tremaine et al. (2002) respect to that for radio-quiet quasars,
which is only partly due to the possible cosmology evolution of
the $\mbh- \sigma_*$ relation and the luminosity bias. (2) The
radio luminosity is correlated to the central SMBH mass and the
Eddington ratio, $ \propto
\mbh^{1.28^{+0.23}_{-0.16}}(\lb/\ledd)^{1.29^{+0.31}_{-0.24}}$ for
radio-quiet quasars and $ \propto
\mbh^{3.10^{+0.60}_{-0.70}}(\lb/\ledd)^{4.18^{+1.40}_{-1.10}}$ for
radio-loud quasars. Weaker correlation coefficient of the radio
luminosity dependence upon the mass and the Eddington ratio for
radio-loud quasars shows other physical effects would account for
their radio luminosity, such as the SMBH spin.

\section*{ACKNOWLEDGMENTS}
We thank Luis C. Ho for his very helpful comments. We thank Dr. M.
Wu and Z. H. Fan for the discussion. We thank the anonymous
referee for his/her comments and instructive suggestions. This
work has been supported by the NSFC (Nos. 10733010, 10403005,
10325313, 10233030 and 10521001), the Science-Technology Key
Foundation from Education Department of P. R. China (No. 206053),
and China Postdoctoral Science Foundation (No. 20060400502). Chen
Hu thanks T. A. Boroson \& M. Vestergaard for the I ZW 1 iron
templates and their so kindly suggestions on the spectral fitting.

Funding for the creation and distribution of the SDSS Archive has
been provided by the Alfred P. Sloan Foundation, the Participating
Institutions, the National Aeronautics and Space Administration,
the National Science Foundation, the US Department of Energy, the
Japanese Monbukagakusho, and the Max Planck Society. The SDSSWeb
site is http://www.sdss.org. The SDSS is managed by the
Astrophysical Research Consortium for the Participating
Institutions. The Participating Institutions are the University of
Chicago, Fermilab, the Institute for Advanced Study, the Japan
Participation Group, The Johns Hopkins University, the Korean
Scientist Group, Los Alamos National Laboratory, the Max Planck
Institute for Astronomy, the Max Planck Institute for
Astrophysics, New Mexico State University, the University of
Pittsburgh, the University of Portsmouth, Princeton University,
the United States Naval Observatory, and the University of
Washington.

This research has made use of the NED database, which is operated
by the Jet Propulsion Laboratory, California Institute of
Technology, under contract with the National Aeronautics and Space
Administration.


\end{document}